\documentstyle[12pt]{article}
\begin{document}
\baselineskip .3in
\begin{titlepage}
\begin{center}{\large{\bf The colours of quarks as new degrees of freedom}}
\vskip .2in S.Mukherjee(Banerjee)$^{\dag}$ \\
 \vskip .1in Dublin
Institute of Technology,Dublin,Ireland. \vskip .1in
S. N. Banerjee  \\
\vskip .1in
Department of Physics, Jadavpur University \\
Calcutta 700032, India.\\

\end{center}

\vskip .3in \noindent {\bf Abstract} \vskip .1in \noindent The
origin of the colours of quarks has been explored and the number
of colours equal to three has been derived from the fractal
properties suggested in the statistical model.The quark gluon
coupling constant has been reproduced and the properties of the
intrinsic electric charges of quarks have also been studied.


\vskip .3in

\noindent $^{\dag}$ E-mail:sarbarimukherjee09@gmail.com

\end{titlepage}

\newpage
 \vskip .1in \noindent
{\bf 1.Introduction} \vskip .1in \noindent It is now accepted that
quarks fundamental constituents of matter,possess a new and
hitherto unknown internal degrees of freedom called colour,to
resolve several difficulties in the quark model.The colour quantum
number is hidden as all known hadrons are colour singlets and it
is visible only when it is probed at a momentum transfer which
allows us to resolve the individual quark constituents.\vskip .1in
\noindent The statistical model of the hadron as a quarkonium
system,proposed almost three decades ago[1],has been widely used
to study the properties of hadrons.The model,however,has undergone
modifications along with its various ramifications and
applications since its very inception[2].The power-law scaling,the
fractal properties as well as the universality of hadrons have
been analysed recently in the framework of the model[3].The
scaling exponent in the production formula of hadrons suggested by
the model,has been found to be in conformity with the recent
experimental findings.The colour factor ratio[4] has been derived
recently from the fractal properties of hadrons and is found to be
in exact agreement with the corresponding QCD value.In the current
investigation we have attempted to trace the origin of the colour
of quarks in the framework of the model.The real and virtual
quarks in the sea of quark-antiquark ($q\bar q$) pairs in the
model have been assumed to be identical and indistinguishable from
one another and we arrive at the number density of quarks
$n_{q}(r)$ inside the hadron.This has led us to derive the number
of colours $N_{c}$ of quarks equal to 3,as new degrees of
freedom.As a result, the quark gluon coupling constant has also
been derived and is found to be in exact agreement with the
corresponding QCD estimate. \vskip .1in \noindent Using the
theoretically derived value of $N_{c}=3$ as an input, it has been
possible to explore also the origin of the fractional electric
charges of quarks of different flavours and to obtain the bounds
of their absolute magnitudes. On the other hand,it was suggested
earlier[5] that the charges of the $u$ and $d$ quarks have to be
known from some external source in order to infer the numerical
value of $N_{c}$. \vskip .1in \noindent {\bf 2. Colours as new
degrees of freedom} \vskip .1in \noindent  \noindent Due to the
identity and indistinguishability of the real and virtual quarks
in the sea as described in the model, a special kind of geometric
complexity emerges and the hadron turns out to be a fractal
object.The local density of quarks(or anti-quarks)has been found
to depend only on the size parameter $r_{0}$ of the hadron as
[1-4] \vskip .1in \noindent
\begin{equation}
n_{q}(r)= f(r_{o})g(\xi)
\end{equation}
for $r\leq r_{0}$ and $n_{q}(r)=4$ for $r\geq r_{0}$. Here
$\xi=r_{0}-r$, $f(r_{0})=\frac{315}{64{\pi}r_{0}^{9/2}}$ and
$g(\xi)=\xi^{d}$ where $d=3/2$. The fractal curve represented by
$g(\xi)$ correspond to the region bounded by the interface and
using the polygonal approximation to cover the fractal,we come
across the minimum number of line segments,each of measure
$\epsilon^{d}$ with a small interval of length $\epsilon$.Thus
$g(\xi)$ represents a d-dimensional curve embedded into the
$D=9/2$ dimensional space and $d=3/2$ appearing as an exponent in
defining the measure.Again,the spin degeneracy factor,arising from
the two allowed values of the spin quantum number is 2 as the
quark is a spin 1/2 particle.Consequently, the resulting factor
turns out to be equal to 2d=3.This new factor appears as the
internal degrees of freedom to be assigned to the quark and has
been named as the colour degrees of freedom i.e. $N_{c}=2d=3$.
\vskip .1in \noindent {\bf 3. Applications} \vskip .1in \noindent
The colour factors $C_{F}$ and $C_{A}$ are known to determine the
intensity of gluon radiation off a quark and a gluon respectively
and in the case of $SU(3)$ for QCD[6,7,8], one comes across the
probability for a gluon to split into two gluons to be
proportional to $C_{A}=N_{c}=3$ corresponding to the three
colours[5].On the other hand,using the numerical value of
$C_{A}/C_{F}$ derived previously by us in the framework of the
model [4] in conjunction with the derived value of $N_{c}=3$ in
the current work,we get directly the value of $C_{F}=4/3$, the
quark gluon coupling constant and it agrees with the corresponding
QCD estimate. \vskip .1in \noindent The Drell ratio $ R\equiv
\sigma(e^{+}e^{-}\rightarrow hadrons)/4{\pi}{\alpha}^{2}/s $ where
$\alpha =e^{2}/4{\pi}$[6] can now be recast directly as
\begin{equation}
R=3 \Sigma_{_{i=1}}^{N_{f}}e_{i}^{2}
\end{equation}
\vskip .1in \noindent Here $e_{i}$ is the intrinsic electric
charge of the ith quark and $N_{f}$,the number of quark flavours
which may contribute to the process and is constrained by
$2m_{N_{f}}<\sqrt{s}$ where $s$ is the centre of mass total energy
squared of the $e^{+}e^{-}$ system and $m_{{N}_{f}}$ is the mass
of the quark of flavour $f$.For the energy region $\sqrt{s}< 3$
Gev,only the u,d and s quarks contribute since this region is
below the c-quark production threshold.The value $R\simeq 2$ is
apparent below the threshold for producing charmed particles at
energy $\simeq 3.7 Gev[6]$. \vskip .1in \noindent Therefore,
\vskip .1in \noindent $R=\frac{2N_{c}}{3}=2$ we get
\begin{equation}
2=3(e_{u}^{2}+e_{d}^{2}+e_{s}^{2})
\end{equation}
for $E_{cm}> 2Gev > 2m_{u}$, $2m_{d}$, $2m_{s}$[7]. Hence
\begin{equation}
\frac{2}{3}= e_{u}^{2}+e_{d}^{2}+e_{s}^{2}
\end{equation}
Therefore,without any further input of experimental data,we get
the sum of the squares of the electric charges of quarks of
flavours u,d and s,equal to 2/3.Each term on the right hand side
of (4) is therefore a proper fraction.In other words,the absolute
magnitude of the charge of each quark corresponding to the three
flavours is less than 1.Using $N_{c}^{2}=9$ corresponding to our
derived value of $N_{c}=3$ the decay width $\Gamma$ for the
$\pi^{^{0}}\rightarrow 2\gamma $ process may now be directly
recast as
\begin{equation}
\Gamma =\frac {9({e_{u}}^{2}-{e_{d}}^{2})^{2}{\alpha}^{2}\cdot
m_{\pi}^{3}}{64\pi^{2}f_{{\pi}^{2}}}
\end{equation}
where $f_{\pi}= 91 Mev,$ the pion decay constant for
$\pi\rightarrow \mu\nu $ decay.Using the experimental value of
$\Gamma \simeq 7.57 ev $ as the input,we get
\begin{equation}
{e_{u}}^{2}-{e_{d}}^{2}\simeq 1/3
\end{equation}
Combining (5) and (6),we get
\begin{equation}
e_{s}^{2}=1-2e_{u}^{2}
\end{equation}
\begin{equation}
e_{s}^{2}< 1 \indent \mbox{or,} \mid e_{s}\mid <1
\end{equation}
Hence \begin{equation} 2e_{u}^{2}<1 \indent \mbox{i.e.}\indent
e_{u}^{2}<1/2\indent \mbox{ or,}\indent \mid e_{s}\mid <1/\sqrt{2}
\end{equation}
\vskip .1in \indent Therefore,we get from(6),
\begin{equation}
e_{d}^{2}<1/6 \indent \mbox{or,} \mid e_{d}\mid < 1/\sqrt{6}
\end{equation}
\vskip .1in \noindent These bounds on the intrinsic electric
charges of quarks are obtained without any further experimental
data as inputs.For the region $4Gev < \sqrt{s}< 9 Gev$,there is an
additional contribution from the c quark and the corresponding
experimental value of $R\simeq 10/3$.Again, above the threshold
for all five quark flavours i.e. for $ \sqrt{s}> 10 Gev $,
$R_{exp}\simeq 1/3$.Using these two observed values of R for the
two regions,in conjunction with(2),we arrive at $e_{c}^{2}=4/9$
i.e. $|e_{c}|=2/3$ and $e_{b}^{2}=1/9$ i.e.$|e_{b}|=1/3$
.Therefore,using $N_{c}=9$,the fractional nature of the intrinsic
electric charges of quarks follows as natural consequence and the
absolute magnitudes of the charges of c and b quarks are
explicitly determined. \vskip .1in \noindent {\bf 4.Dimensions and
conclusions} \vskip .1in \noindent The origin of the colour degree
of freedom of quarks is attributed to the fractal properties
assigned by the model.The number of colours $N_{c}$ has been
derived theoretically to be equal to 3 through its dependence on
$d=3/2$ and the spin projection.Further,interesting bounds on the
absolute magnitude of the intrinsic charges are obtained,which in
turn suggest their fractional values.
\newpage
\vskip .2in \noindent {\bf References}

\noindent[1]S.N.Banerjee,A.K.Sarkar,Had.J.4(1981)2003;5(1982)2157;6(1983)440;
S.N.Banerjee,J.Phys.G8(1982)L61; S.N.Banerjee and
A.Chakroborty,Ann.Phys.(N.Y.)
150(1983)150;
Int.J.Mod.Phys.A16(2001)201;17(2002)4939.

\noindent [2]S.N. Banerjee,N.N.Begum and
A.K.Sarkar,Had.J.11(1988)243;12(1989)179;13(1990)75.

\noindent [3] S.N.Banerjee and S.Banerjee,Phys.Lett.B644(2007)45.

\noindent [4]S.Mukherjee and S.N.Banerjee,Mod.Phys.Lett
A,24(2009),No.7,509.

\noindent [5]Quantum Chromodynamics,G.Dissertori,I.Knowles and
M.Schmelling,Clarendon Press,Oxford(2003)23,17,16.

\noindent [6] Foundations of Quantum Chromodynamics,T.muta,world
Scientific,Singapore(1998)11.

\noindent [7] Quantum Chromodynamics,W.Greiner,S.Schram and
E.Stein-springer Verlag-Berlin(2007)166.

\end{document}